\documentclass[aps,prd,amsmath,amssymb]{revtex4}

\usepackage{graphicx}
\usepackage{dcolumn}
\usepackage{bm}
\usepackage{amsmath}
\usepackage{epstopdf}
\usepackage{amsfonts}
\usepackage{amssymb}
\usepackage{epsfig}
\usepackage{tabularx}
\usepackage{color}
\usepackage{soul}

\usepackage[colorlinks=true,citecolor=blue,urlcolor=magenta,breaklinks]{hyperref}

\newcommand{\be}{\begin{equation}}
\newcommand{\en}{\end{equation}}
\newcommand{\bea}{\begin{eqnarray}}
\newcommand{\ena}{\end{eqnarray}}

\begin{document}

\title{Condensate Dark Stars beyond the Mean-Field Approximation: The Lee-Huang-Yang correction}

\author{
G.~Panotopoulos,
\footnote{
\href{mailto:grigorios.panotopoulos@ufrontera.cl}{grigorios.panotopoulos@ufrontera.cl}
}
}

\address{
Departamento de Ciencias F{\'i}sicas, Universidad de la Frontera, Avenida Francisco Salazar, 01145 Temuco, Chile. 
}

\begin{abstract}
We study structural properties of self-gravitating fluid spheres made of a dilute, homogeneous and ultracold Bose gas assuming repulsive, short-range interactions. For the first time we include the Lee-Huang-Yang correction to the usual polytropic equation-of-state of index $n=1$, which goes beyond the Hartree mean-field approximation taking into account quantum fluctuations. We find that the correction has a considerable impact on the M-R relationships and other properties of condensate dark stars, such as factor of compactness and tidal Love numbers. The impact is more significant for equation-of-states that support larger highest stellar masses.
\end{abstract}

\maketitle

\section{Introduction}

Since the pioneer works of both F.~Zwicky in the 30’s and V.~Rubin in the 70's, we are convinced that most of the non-relativistic matter in the Universe is dark, usually referred to as cold dark matter (DM). 
Zwicky analyzed the dynamics of the Coma galaxy cluster\cite{Zwicky}, while a few decades after that the observations made by Rubin helped in determining the galaxy rotation curves \cite{Rubin}. In modern times, well
established current data, coming from different cosmological and astronomical sources, seem to confirm the existence of DM \cite{Turner}, despite the fact that its nature and origin still remain a mystery. For a review on DM see \cite{Olive, Munoz}, and for more recent reviews
on DM detection searches see \cite{Gascon, Gaskins, Kahl}.

\smallskip

The latest cosmological data from the PLANCK collaboration \cite{Planck:2018vyg} suggest that DM is the dominant component of the non-relativistic matter in the Universe. Despite the success of the standard (concordance) cosmological model at large cosmological scales (1~Mpc), based on the cold dark matter paradigm and a tiny positive cosmological constant, a number of shortcomings at galactic and sub-galactic scales (a few kpc) persists, such as the core/cusp problem and the missing satellite problem. For reviews on the DM crisis at small scales see e.g. \cite{Suarez:2013iw, Li:2013nal}. Those problems may be tackled including self-interactions \cite{Spergel:1999mh, Dave:2000ar}, as any cuspy feature will be smoothed out by DM collisions, or if DM consists of ultralight scalar particles with a tiny mass $m \ll eV$ \cite{Hu:2000ke, Hui:2016ltb}. In those models, ultralight bosons can cluster forming macroscopic Bose-Einstein condensates at solar masses or even higher masses. In the case of spinless bosons those self-gravitating clumps are called scalar boson stars \cite{Kaup:1968zz, Ruffini:1969qy, Khlopov:1985fch, Colpi:1986ye, Tkachev:1986tr, Tkachev:1991ka, Kusmartsev:1990cr, Schunck:2003kk, Souza:2014sgy, Pires:2012yr, Eby:2015hsq, Croon:2018ybs}.

\smallskip

It was S.~N.~Bose in the 20's who first pointed out \cite{Bose}, and shortly after that expanded by A.~Einstein \cite{E1, E2}, that in a quantum gas made of bosons, indistinguishability of the particles requires a new statistical description, now known as Bose-Einstein statistics. If the temperature of the gas is low enough a new exotic form of matter is formed. Bose-Einstein-Condensate (BEC) in three space dimensions is exclusively driven by the quantum statistics of the bosons, and not by the interactions between them. The BEC, considered to be the fifth state of matter (after gases, liquids, solids and plasma), is manifested in the classical example of Helium-4 superfluidity \cite{Kapitza, Allen}, and led to the Nobel Prize in Physics in 2001 \cite{url}. DM may be a Bose-Einstein condensate and thus it might solve the cusp/core problem \cite{Boehmer:2007um, Harko:2011xw}. Moreover, it is possible that condensate dark stars just might exist \cite{Li:2012sg}, see also \cite{Chavanis:2011cz} on BEC general relativistic stars.

\smallskip

The gravitational wave signal emitted during the inspiral and subsequent collisions of two astronomical objects in a binary system contains a wealth of information regarding the nature of the colliding stars. The imprint of the equation-of-state (EoS) within the signals emitted during binary coalescences is mainly determined by adiabatic tidal interactions that are characterized by a set of coefficients, known as the tidal Love numbers, $k$, and the corresponding deformabilities, both dimensionful, $\lambda$, and dimensionless, $\Lambda$. It was Love in the beginning of the 20th century who first introduced tidal deformability in Newtonian gravity \cite{Love1,Love2}, as he was trying to understand the tidal response of the Earth. In the case of objects with spherical symmetry, Love introduced two dimensionless numbers to describe tidal response and deformability. In particular, the first number, $h$, describes the relative deformation of the body in the longitudinal direction (with respect to the perturbation), while the other one, $k$, describes the relative deformation of the gravitational potential. In modern times the consideration of self-gravitating compact objects requires a relativistic theory of tidal deformability, which was developed almost 20 years ago \cite{flanagan,hinderer,damour,Lattimer,poisson} for static, spherically symmetric neutron stars, strange quark stars and black holes. Naturally, the key deformability parameter is now the relativistic generalization of $k$, since in metric theories of gravity the role of the gravitational potential is now played by the metric tensor.

\smallskip

It is well-known that structural properties of stars, such
as mass and radius, depend crucially on the EoS, which unfortunately is poorly known. In particular, as far as studies on condensate dark stars are concerned, as of today the usual polytropic equation of index $n=1$ has been employed within the mean-field approximation ignoring quantum fluctuations. Beyond the mean-field approximation, the Lee-Huang-Yang (LHY) correction \cite{Lee:1957zza, Lee:1957zzb} takes into account quantum fluctuations around the mean-field value. Although in condensed matter physics the LHY correction is known, to the best of our knowledge, no one has investigated its effect on properties of condensed dark stars. It is the goal of the present article to study for the first time the impact of the LHY correction on structural properties of condensate dark stars, thus filling a gap in the literature. We anticipate that we find significant deviations from the standard results reported in \cite{Chavanis:2011cz}, see the discussion in Section 4, figures and conclusions.

\smallskip

Our work is organized as follows: After this introductory section, we summarize the structure equations and tidal Love numbers for fluid spheres within GR in Section 2, while in the third section we review the Gross-Pitaevskii-Poisson system and the EoS within the Hartree mean-field approximation. Next, in Section 4 we include the LHY correction and we integrate the TOV as well as the Riccati equations numerically to compute the stellar mass and radius as well as the factor of compactness and the tidal Love numbers. The Harrison-Zeldovich-Novikov criterion for stability is discussed as well. We summarize our work in the fifth section with some concluding remarks. Finally, for self-completeness we have included an Appendix after the Conclusion Section. Moreover, throughout the manuscript we adopt the mostly positive metric signature $\{-,+,+,+\}$, and we work in natural geometric units in which the physical constants $G, c, k_B, \hbar$ are set to unity, $G=c=k_B=\hbar=1$.

\section{Structural Properties of Fluid Spheres within General Relativity}

\subsection{TOV Equations for Interior Solutions}

We work with fluid spheres made of isotropic matter in (1+3)-dimensional geometries, with a vanishing cosmological constant. Although rotation could be present in condensate dark stars, in the present work we shall consider non-rotating objects only. The inclusion of a non-vanishing angular momentum, at least in the slow-rotating case, may be implemented afterwards in a straightforward manner, see e.g. \cite{Staykov:2014mwa, Panotopoulos:2021dtu}.

We shall briefly review the structure equations for interior stellar solutions, starting from the field equations of Einstein's General Relativity \cite{Einstein:1915ca}
\begin{equation}
    G_{mn} \equiv R_{mn} - \frac{1}{2} \: R \: g_{mn} = 8 \pi  T_{mn},
\end{equation}

\medskip

\noindent where $T_{mn}$ is the energy-momentum tensor of matter content, $g_{mn}$ is the metric tensor, $R_{mn}$ and $R$ are the Ricci tensor and Ricci scalar, respectively, while $G_{mn}$ is the Einstein tensor.

\smallskip

For static, spherically symmetric geometries the most general line element in Schwarzschild-like coordinates $\{ t, r, \theta, \phi \}$ is given by
\begin{equation}
    d s^2 = - e^{\nu} d t^2 + e^{\lambda} d r^2 + r^2 (d \theta^2 + \sin^2 \theta \: d \phi^2) , 
\end{equation}

\medskip

\noindent while the isotropic matter content, viewed as a perfect fluid, is described by a stress-energy tensor of the following form
\begin{equation}
T_a^b = Diag(-\rho, p, p, p),
\end{equation}

\medskip

\noindent where $p$ is the pressure of the fluid, in both radial and tangential directions, while $\rho$ is its energy density. What is more, depending on the matter content, $p$ and $\rho$ satisfy a certain EoS $F(p, \rho)=0$.

\smallskip

To compute the stellar mass, $M$, and radius, $R$, we need to integrate the Tolman-Oppenheimer-Volkoff (TOV) equations \cite{Oppenheimer:1939ne, Tolman:1939jz}. Those are the structure equations that permit us to obtain stellar interior solutions describing hydrostatic equilibrium. The TOV equations, assuming only one fluid component, are given by
\begin{align}
    \displaystyle m'(r) &= 4 \pi r^2 \rho (r) , \\ 
    \displaystyle \nu' (r) &= 2 \: \frac{m(r) + 4 \pi r^3 p(r)}{ r^2 \left( 1 - 2 m(r) / r \right) } , \\
    \displaystyle p'(r) &= - [ \rho(r) + p(r) ] \; \frac{\nu' (r)}{2} ,
\end{align}

\smallskip

\noindent where the prime denotes differentiation with respect to the radial coordinate $r$, while the mass function, $m(r)$, is defined as usual by
\begin{equation}
    \displaystyle e^{\lambda} = \frac{1}{1 - \frac{2 \: m(r)}{r}} .
    \label{eq:2}
\end{equation}

\medskip

In practice we only need to solve a system of two coupled differential equations
\begin{align}
    \displaystyle m'(r) &= 4 \pi r^2 \rho (r) , \\ 
    \displaystyle p'(r) &= - [ \rho(r) + p(r) ] \;  \frac{m(r) + 4 \pi r^3 p(r)}{ r^2 \left( 1 - 2 m(r) / r \right) } .
\end{align}

\smallskip

To find the solution, we integrate the structure equations throughout the star imposing initial conditions at the center ($r = 0$), and then we impose the matching conditions at the surface of the star ($r = R$). In particular, the initial conditions at the origin are as follows 
\begin{equation}
    m(0) = 0 ,
\end{equation}
\begin{equation}
    p(0) = p_c ,
\end{equation}

\medskip

\noindent with $p_{c}$ being the central pressure, while the exterior vacuum solution is given by the well-known Schwarzschild geometry \cite{Schwarzschild:1916uq}
\begin{equation}
d s^2 = -f(r) d t^2 + f(r)^{-1} dr^2 + r^2 (d \theta^2 + \sin^2 \theta \: d \phi^2), \; \; \; \; \; \; \; f(r) = 1-2 \frac{M}{R} .
\end{equation}

Furthermore, the matching conditions yield
\begin{equation}
p(R) = 0,
\end{equation}
\begin{equation}
m(R) = M, 
\end{equation}
\begin{equation}
e^{\nu(R)} = 1-2 \frac{M}{R} .
\end{equation}

\medskip

The first two conditions allow us to compute the radius and the mass of the star. Finally, the other metric potential, $\nu(r)$, may be computed by
\begin{equation}
    \displaystyle \nu (r) = \ln \left( 1 - \frac{2 M}{R} \right) + 2 \int_R^r \frac{m(x) + 4 \pi x^3 p(x)}{ x^2 \left( 1 - 2 m(x) / x \right) } \: dx .
    \label{eq:3}
\end{equation}
where we have used the third matching condition.

\subsection{Gravito-electric Tidal Love Numbers}

For a complete description of the relativistic theory of tidal Love numbers for compact objects the interested reader may consult for instance \cite{flanagan,hinderer,damour,Lattimer,poisson}. 

\smallskip

Let us consider a binary system made of the principal star and its companion object. The former is subjected to an external gravitational field, $\Phi_{ext}$, produced by the companion star in the binary. The star under discussion reacts to the external field by deforming and thus developing a quadrupolar moment $Q_{ij}$. The tidal deformability, $\lambda$, is defined by the following linear response 
\begin{equation}
Q_{ij} = - \lambda \: E_{ij},
\end{equation}
where the quadrupolar moment, $Q_{ij}$, is given by
\begin{equation}
Q_{ij} = \int d^3x \delta \rho(\vec{x}) \: (x_i x_j - \frac{1}{3} r^2 \delta_{ij}),
\end{equation}
while the tidal field, $E_{ij}$, is given by
\begin{equation}
E_{ij} = \frac{\partial^2 \Phi_{ext}}{\partial x^i \partial x^j},
\end{equation}
and the spatial indices take three values $i,j=1,2,3$.

\smallskip

The tidal Love number $k$, a quadrupole moment number and dimensionless coefficient, depends on the inner structure of the star and thus on its underlying EoS. It is directly related to two auxiliary quantities commonly referred to as "deformabilities", denoted $\lambda$ (dimensionful) and $\Lambda$ (dimensionless), which are defined as follows:
\begin{align}
\lambda &\equiv \frac{2}{3} k R^5,
\label{eq:Love1}
\\
\Lambda &\equiv \frac{2 k}{3 C^5},
\label{eq:Love2}
\end{align}
where $C=M/R$ is the usual factor of compactness of the star. 

\smallskip

Next, the tidal Love number can be computed in terms of two numbers only, namely $C$ and $y_R$ as follows \cite{flanagan,hinderer,damour,Lattimer,poisson}
\begin{align}
k &= \frac{8C^5}{5} \: \frac{K_{o}}{3  \:K_{o} \: \ln(1-2C) + P_5(C)} ,
\label{elove}
\\
K_{o} &= (1-2C)^2 \: [2 C (y_R-1)-y_R+2] ,
\\
y_R &\equiv y(r=R) ,
\end{align}
where $P_5(C)$ is a fifth-order polynomial that is found to be
\begin{align}
\begin{split}
P_5(C) = \: & 2 C \: \Bigl( 4 C^4 (y_R+1) + 2 C^3 (3 y_R-2) \ + 
\\
&
2 C^2 (13-11 y_R) + 3 C (5 y_R-8) -
\\
&
3 y_R + 6 \Bigl)  ,
\end{split}
\end{align}
%
%
while the function $y(r)$ satisfies the Riccati differential equation \cite{Lattimer}:
\begin{align}
\begin{split}
r y'(r) + y(r)^2 &+ y(r) e^{\lambda (r)} \Bigl[1 + 4 \pi r^2 ( p(r) - \rho (r) ) \Bigl] 
\\
&+ r^2 Q(r) = 0 ,
\end{split}
\end{align}
supplemented by the initial condition at the center, $r \rightarrow 0$, $y(0)=2$, where the function $Q(r)$, not to be confused with the tensor $Q_{ij}$, is computed to be
\begin{align}
\begin{split}
  \displaystyle Q(r) = 4 \pi e^{\lambda (r)} \Bigg[ 
  5 \rho (r) 
  &+ 9 p(r) + \frac{\rho (r) + p(r)}{c^2_s(r)} 
  \Bigg] 
  \\
  &- 6 \frac{e^{\lambda (r)}}{r^2} - \Bigl[\nu' (r)\Bigl]^2 .
\end{split}
\end{align}
with $c_s^2 \equiv dp/d\rho$ being the speed of sound. 

\smallskip

We remark in passing that since $k \propto (1-2C)^2$, tidal Love numbers of black holes vanish due to the fact that $C=1/2$ in the case of Schwarzschild black holes within Einstein's gravity. This is an intriguing result of classical GR saying that tidal Love numbers of black holes, as opposed to other types of compact objects, are precisely zero. Therefore, a measurement of a non-vanishing $k$ will be a smoking-gun deviation from the standard black hole of Einstein's theory. Moreover, regarding gravity waves observatories, the Einstein Telescope will pin down very precisely the EoS of neutron stars \cite{Iacovelli:2023nbv}, while the Laser Interferometer Space Antenna is able to probe even extremely compact objects (with a factor of compactness $C > 1/3$), and it will set constraints on the tidal Love numbers of highly-spinning central objects at $\sim 0.001-0.01$ level \cite{Piovano:2022ojl}.

\section{Bose-Einstein Condensate: Mean-Field Approximation}

Since not all the readers are familiar with the details of the Bose-Einstein condensate and how to derive the underlying equation-of-state for condensate dark stars, in this Section we shall discuss in detail how to obtain the polytrope $p=K \rho^2$ widely used in the literature. We have also included 2 Appendices in the end of the article for self-completeness.

\smallskip

We shall be considering a dilute and ultracold gas of identical 
spin-zero bosons below the critical temperature of the BEC, $0 < T < T_c$, where $T_c$ is given by \cite{PhD}
\begin{equation}
T_c = \frac{2 \pi}{m} \left( \frac{n}{\zeta(3/2)} \right)^{2/3} \approx 3.31 \frac{n^{2/3}}{m}
\end{equation}
where $\zeta(3/2) \approx 2.612$ is Riemann’s zeta function, with $m$ being the mass of the particles in the system, and $n$ being its number density given by
\begin{equation}
n = \int \frac{d^3\vec{k}}{(2 \pi)^3} n_B(\xi_k)
\end{equation}
where $n_B$ is the Bose-Einstein distribution
\begin{equation}
n_B(\xi_k) = \frac{1}{e^{\beta (\epsilon_k-\mu)}-1},
\end{equation}
with $\mu$ being the chemical potential, $\beta=1/T$, and $\epsilon_k=k^2/2m$ being the energy of each particle with $k$ being the wave-number.

\smallskip

The condition for BEC is that the chemical potential vanishes at $T_c$ and at lower temperatures
\begin{equation}
\mu(T) = 0, \; \; \; \; \; \; T \leq T_c
\end{equation}
The reason is that the wave-function of the BEC can accommodate a large number of particles, and we may add one more particle to the system at no energy cost. 

\smallskip

In a dilute and ultra cold boson gas the cross-section of the two body elastic scattering at low energies is given by
\begin{equation}
\sigma = 4 \pi a_s^2
\end{equation}
with $a_s$ being the scattering length. This may be easily shown making use of the expression for the elastic cross-section in the partial-wave analysis \cite{QM}
\begin{equation}
\sigma = \frac{4 \pi}{k^2} \sum_{l=0}^\infty sin^2\delta_l
\end{equation}
where $m$ is the mass of particles in the system, $k$ is the wave number, $k=\sqrt{2 m \epsilon}$, $l$ is the angular degree, and
$\delta_l$ are the phase-shifts. At low energies $k a_s \ll 1$ the first term ($l=0$, s-wave scattering) is the dominant one in the series expansion, and therefore
\begin{equation}
\sigma \approx \frac{4 \pi}{k^2} sin^2\delta_0
\end{equation}
where the phase-shift at low energies $\delta_0 \approx k a_s $, and $sin(k a_s) \approx k a_s$. Replacing we obtain $\sigma = 4 \pi a_s^2$.

\smallskip

Since at low energies the details of the two-body interaction potential are not important, we may consider the case of a Dirac delta function
\begin{equation}
V(\vec{r}_1,\vec{r}_2) = g \: \delta^3(\vec{r}_1-\vec{r}_2),  \; \; \; \; \; \; g>0
\end{equation}
with a positive coupling constant $g$, which represents a short-range repulsive interaction. In the first Born approximation the elastic cross-section is given by $4 \pi a_s^2$ if $g$ is chosen to be
\begin{equation}
g = \frac{4 \pi a_s}{m}.
\end{equation}
So, the system is characterized by two parameters only ($m,a_s$), namely the particle mass and the scattering length.

\smallskip

{\bf Gross-Pitaevski-Poisson system:} The wave function $\Psi$  of the BEC can accommodate a macroscopically large number of identical bosons, and in the mean-field approximation, i.e. ignoring quantum fluctuations, it satisfies the time-dependent Gross-Pitaevskii equation \cite{Gross:1961kqh, Pitaevskii, Dalfovo:1999zz}
\begin{equation}
i \frac{\partial \Psi}{\partial t} = \left[ -\frac{\nabla^2}{2 m} + g |\Psi|^2 \right] \Psi 
\end{equation}
also known as the non-linear Schr{\"o}dinger equation. 

One may easily obtain this equation considering the non-relativistic limit of the Klein-Gordon equation of a self-interacting complex scalar field $\phi(x)$. If the scalar self-interaction potential is $V(\phi)$ the wave equation reads
\begin{equation}
\Box{\phi} = \frac{d V}{d \phi}
\end{equation}
where $\Box$ is the D'Alembertian operator. In the case of a massive scalar field with a quartic self-interaction 
\begin{equation}
V(\phi) = \frac{m^2 c^2}{\hbar^2} |\phi|^2 + \frac{1}{4} \lambda |\phi|^4
\end{equation}
where $\lambda$ is the strength of the interaction, the Klein-Gordon equation takes the form (restoring the physical constants $c,\hbar$)
\begin{equation}
\nabla^2 \phi - \frac{1}{c^2} \frac{\partial^2 \phi}{\partial t^2} = \frac{m^2 c^2}{\hbar^2} \phi + \frac{\lambda}{2} |\phi|^2 \phi
\end{equation}
Next, setting $\phi(x) = \chi \: exp(-i \frac{m c^2 t}{\hbar})$, and upon calculating the partial derivatives with respect to time and spatial coordinates, we obtain the following differential equation for $\chi(x)$
\begin{equation}
\nabla^2 \chi - \frac{1}{c^2} \frac{\partial^2 \chi}{\partial t^2} + 2 i \frac{m}{\hbar} \frac{\partial \chi}{\partial t} = \frac{\lambda}{2} |\chi|^2 \chi.
\end{equation}
In the non-relativistic limit, $c \rightarrow \infty$, we ignore the second time derivative, and thus we finally obtain the time-dependent non-linear Schr{\"o}dinger equation for non-relativistic particles
\begin{equation}
i \hbar \frac{\partial \chi}{\partial t} = \left[ -\frac{\hbar^2 \nabla^2}{2 m} + \frac{\lambda \hbar^2}{4 m} |\chi|^2 \right] \chi .
\end{equation}
For a rigorous derivation of the GP equation the interested reader may consult e.g. \cite{formal}.

\smallskip

The normalization condition of the macroscopic wave function $\Psi$ is chosen to be
\begin{equation}
\int d^3 \vec{r} |\Psi|^2 = N
\end{equation}
with $N$ being the number of bosons in the system. Therefore, the total mass of the system, $M$, is given by
\begin{equation}
M = N m = m \int d^3 \vec{r} |\Psi|^2
\end{equation}
which means that the mass density is given by $\rho=m |\Psi|^2$.

\smallskip

If an external potential is also present, the time-independent GP equation reads
\begin{equation}
\mu \Psi = \left[ -\frac{\nabla^2}{2 m} + V_{ext} + g |\Psi|^2 \right] \Psi
\end{equation}
In gravitational contexts, the external potential is given by $V_{ext} = m \Phi$, where $\Phi$ is the gravitational potential, which in the Newtonian limit satisfies the Poisson equation
\begin{equation}
\nabla^2 \Phi = 4 \pi \rho
\end{equation}
where $\rho$ is the mass density.
Therefore, one needs to consider the Gross-Pitaevski-Poisson (GPP) system \cite{Chavanis:2017loo}
\begin{align}
   \mu \Psi &= \left[ -\frac{\nabla^2}{2 m} + m \Phi + g |\Psi|^2 \right] \Psi, \\ 
    \nabla^2 \Phi &=   4 \pi m  |\Psi|^2.
\end{align}
The non-interacting case, $g=0$, corresponds to the cold and fuzzy dark matter of ultralight particles \cite{Hu:2000ke}.

\smallskip

If the interaction energy is much larger than the kinetic term (Thomas-Fermi approximation) an analytic solution may be obtained as follows. The GP equation implies that $m \Phi \approx -g |\Psi|^2$. Upon replacing into the Poisson equation we obtain the Helmholtz equation for the gravitational potential
\begin{equation}
 \nabla^2 \Phi + (4 \pi m^2 / g) \Phi = 0.
\end{equation}
Applying the method of separation of variables, $\Phi=R(r) Y_l^m(\theta,\phi)$, with $Y_l^m$ being the usual spherical harmonics, the radial part satisfies the spherical Bessel equation \cite{Abramowitz}
\begin{equation}
r^2 R'' + 2 r R' + \left( \frac{2 \pi}{K} r^2 - l (l+1) \right) R = 0, \; \; \; \; \; \; l=0,1,2,3,...
\end{equation}

\smallskip

{\bf Polytropic equation-of-state :} Within the mean-field approximation the equation-of-state of a dilute, ultracold boson system takes the form of a polytrope of the form \cite{Li:2012sg, Chavanis:2017loo}
\begin{equation}
p = K \rho^2, \; \; \; \; \; \;  K = \frac{2 \pi a_s}{m^3}
\end{equation}
This may be seen in two different ways. The first one is the following: Considering the GPP system in the Thomas-Fermi approximation, which consists in ignoring the kinetic term in the GP equation, the radial part of the gravitational potential satisfies the spherical Bessel equation, as mentioned before.
In the case of spherical configurations, $l=0$, and the solution is given by the spherical Bessel function of order 0 \cite{Abramowitz}, 
\begin{equation}
\Phi(r) = A j_0(r) = A \frac{sin(r/a)}{(r/a)}, \;  \;  \;  \;  \;  \; a^2 = \frac{K}{2 \pi}
\end{equation}
\begin{equation}
\rho(r) = m |\Psi|^2 = -\frac{m^2}{g} \Phi(r)
\end{equation}
with $A$ being an arbitrary constant of integration. Therefore, the mass function, which is proportional to the gravitational potential, is given by $j_0(r)$ as well.
On the other hand, the structure equations for fluid spheres in the non-relativistic Newtonian limit, assuming a polytropic EoS with $n=1$, imply a density given by $j_0(r)$ with precisely the same $a$ parameter. This is one of the exact analytic solutions to the Lane-Emden equation, see \cite{Chandra, Shapiro:1983du}.

\smallskip

Another way to reach at the same conclusion is applying the Matsubara formalism and the Feynman rules of diagrammatic expansion to compute the self-energy within the Hartree approximation, see Appendix. One finds that $\Sigma = - g n$, or equivalently $\mu = g n$. In the grand canonical ensemble, the Landau potential, $\Omega$, is defined to be \cite{Pathria}
\begin{equation}
\Omega = E - TS - \mu N
\end{equation}
and it is given by $\Omega=-p V$, with $S$ being the entropy and $p$ being the pressure. The thermodynamic relations at $T \rightarrow 0$ are the following \cite{Pathria}
\begin{align}
    n &= \frac{\partial p}{\partial \mu} , \\ 
    u &= \mu n - p .
\end{align}
If $\mu = g n$, it not difficult to show that the pressure, $p$, and the energy density, $u=E/V$, with $V$ being the volume of the system, are found to be
\begin{equation}
p = u = \frac{g n^2}{2}.
\end{equation}
Since within the Thomas-Fermi approximation the kinetic term is negligible, the energy of the system is dominated by the interaction energy.

\section{Beyond the Mean-Field Approximation: Effect of the Lee-Huang-Yang Correction}

In \cite{Lee:1957zzb} the authors extended the theory of Bogoliubov \cite{Bogo}, and computed the next order correction taking into account quantum fluctuations that are ignored within the mean-field approximation. They found the following expressions for the chemical potential, the energy density and the pressure, valid in the dilute system limit, $n a_s^3 \ll 1$:
\begin{eqnarray}
\mu & = & g n \left[ 1+\frac{32}{3 \sqrt{\pi}} a_s^{3/2} \sqrt{n} \right]      \\
\frac{E}{V} & = & \frac{1}{2} g n^2 \left[ 1+\frac{128}{15 \sqrt{\pi}} a_s^{3/2} \sqrt{n} \right]           \\
p & = &  \frac{1}{2} g n^2 \left[ 1+\frac{64}{5 \sqrt{\pi}} a_s^{3/2} \sqrt{n} \right] 
\end{eqnarray}
where the first term corresponds to the mean-field approximation, while the second term is the LHY correction.

\begin{figure*}[ht!]
\centering
\includegraphics[scale=0.85]{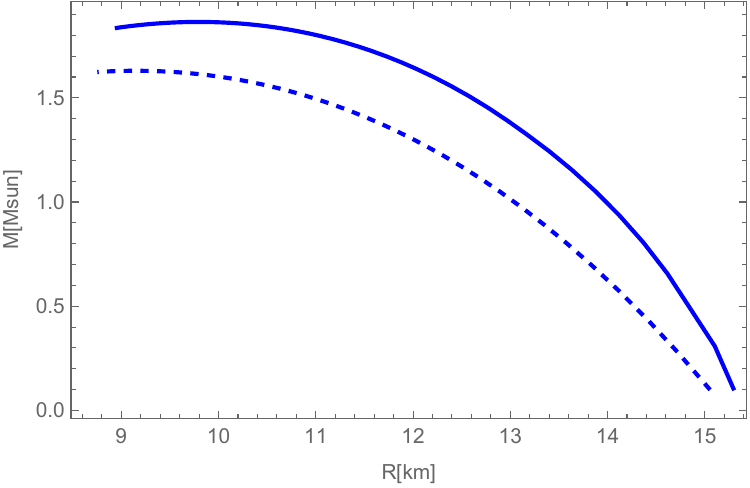} \\
\includegraphics[scale=0.85]{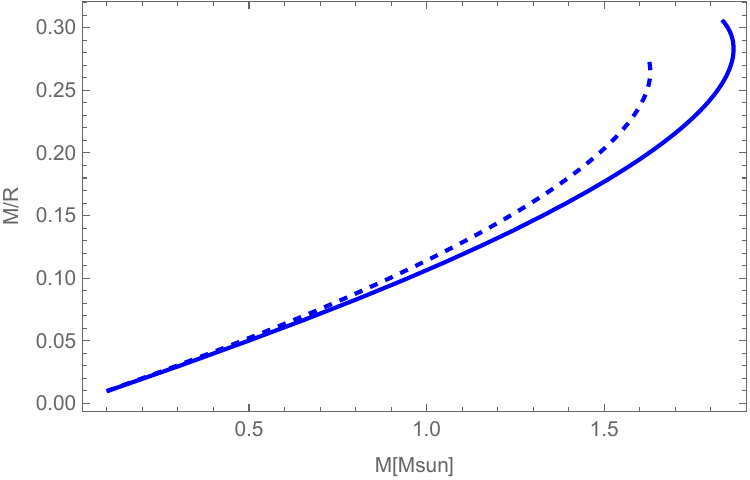}
\caption{
Mass-to-radius relationships (top panel) and factor of compactness versus stellar mass (lower panel) for model A ($m=0.50~GeV, \; a_s=0.10~fm$). Dashed curves correspond to the usual polytrope $p=K \rho^2$, while the solid curves correspond to the EoS including the LHY correction.
}
\label{fig:1} 	
\end{figure*}

\begin{figure*}[ht!]
\centering
\includegraphics[scale=0.9]{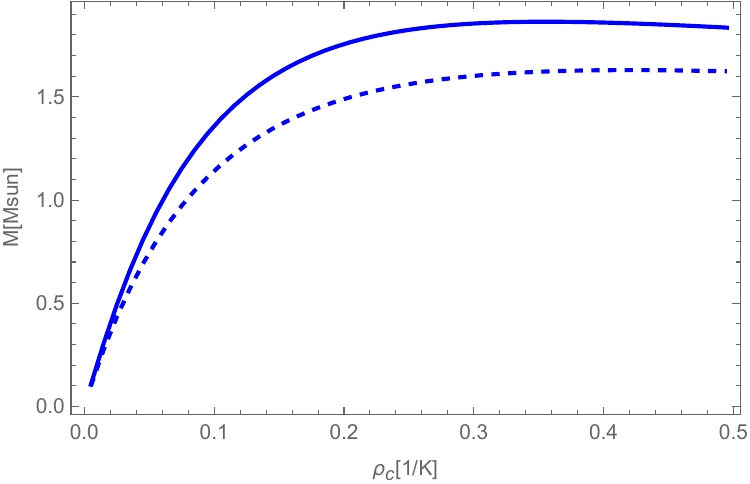} \\
\includegraphics[scale=1.0]{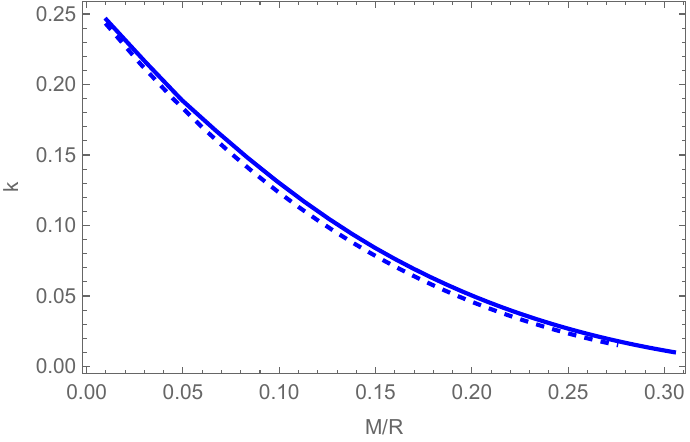}
\caption{
Stellar mass versus central energy density (in units of $1/K$, top panel) and tidal Love numbers versus compactness (lower panel) for model A ($m=0.50~GeV, \; a_s=0.10~fm$). Dashed curves correspond to the usual polytrope $p=K \rho^2$, while the solid curves correspond to the EoS including the LHY correction.
}
\label{fig:2} 	
\end{figure*}

\begin{figure*}[ht!]
\centering
\includegraphics[scale=0.85]{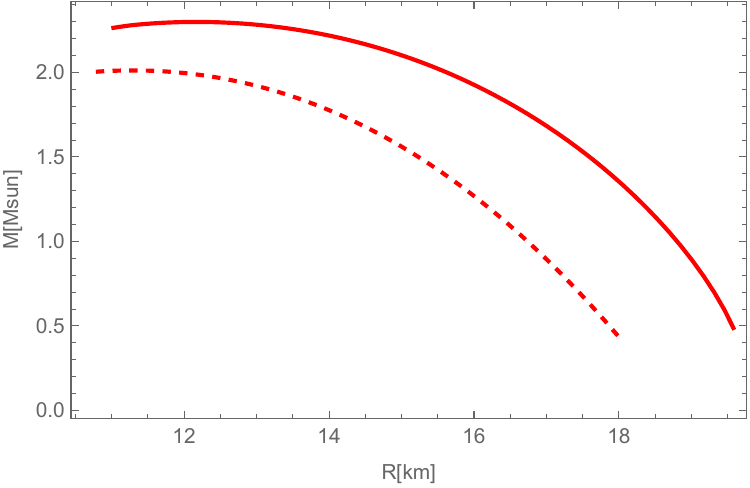} \\
\includegraphics[scale=0.85]{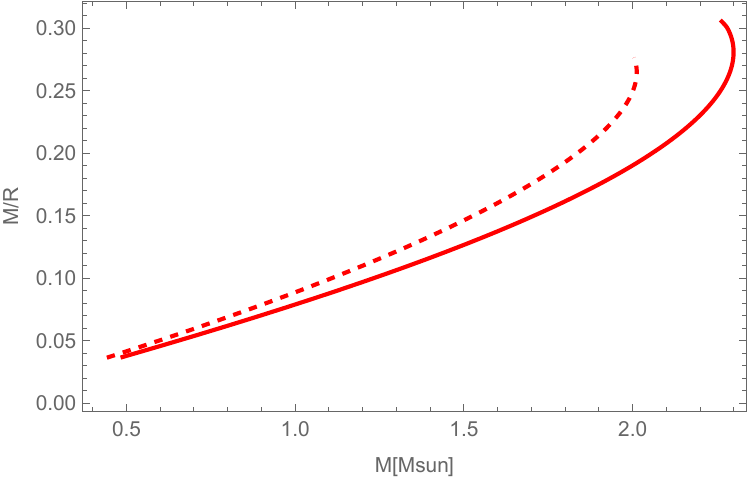}
\caption{
Mass-to-radius relationships (top panel) and factor of compactness versus stellar mass (lower panel) for model B ($m=0.45~GeV, \; a_s=0.11~fm$). Dashed curves correspond to the usual polytrope $p=K \rho^2$, while the solid curves correspond to the EoS including the LHY correction.
}
\label{fig:3} 	
\end{figure*}

\begin{figure*}[ht!]
\centering
\includegraphics[scale=0.95]{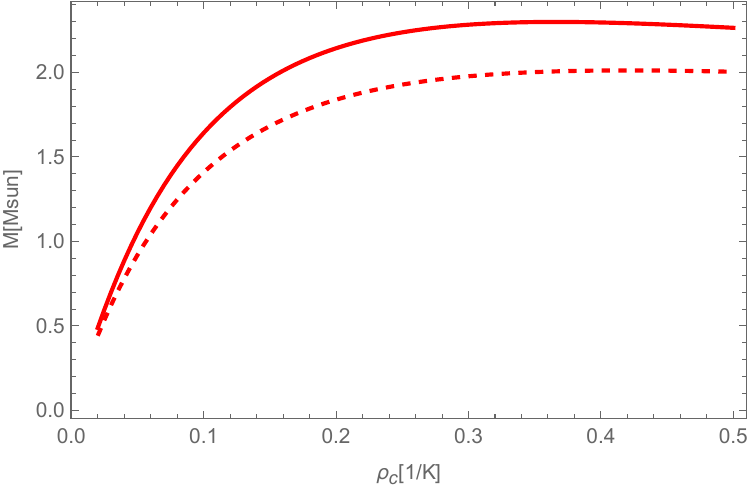} \\
\includegraphics[scale=1.1]{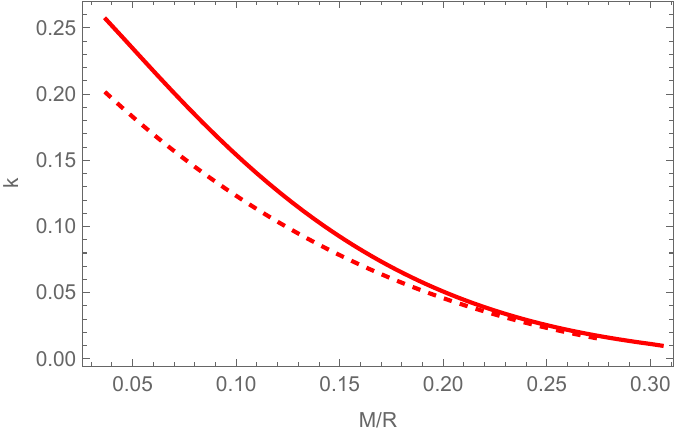}
\caption{
Stellar mass versus central energy density (in units of $1/K$, top panel) and tidal Love numbers versus compactness (lower panel) for model B ($m=0.45~GeV, \; a_s=0.11~fm$). The dashed curve corresponds to the usual polytrope $p=K \rho^2$, while the solid curve corresponds to the EoS including the LHY correction.
}
\label{fig:4} 	
\end{figure*}

In the discussion to follow we shall consider two models as follows:
\begin{equation}
m=0.50 GeV, \; \; \; \;a_s=0.10 fm, \; \; \; \; \textrm{Model A}
\end{equation}
\begin{equation}
m=0.45 GeV, \; \; \; \;a_s=0.11 fm, \; \; \; \; \textrm{Model B}
\end{equation}
For those values the equation-of-state may support stars of mass $(1-2) ~M_{\odot}$ and radius $(10-20)~km$.
In Figures \ref{fig:1}-\ref{fig:4}  below we show the mass-to-radius relationships, the factor of compactness, $C=M/R$, versus stellar mass, the stellar mass versus central energy density, and finally the tidal Love number, $k$, versus the factor of compactness. The curves in blue correspond to model A, while the curves in red correspond to model B. Furthermore, for the sake of comparison, we also show in the same plot the curves corresponding to the usual polytrope with $n=1$ (dashed curves, standard results of \cite{Chavanis:2011cz}), whereas solid lines are for the full EoS including the LHY correction. 

\smallskip

According to our numerical results, model B can accommodate more massive stars. What is more, the factor of compactness remains smaller than the Buchdahl limit, $C < 4/9\approx 0.444$ \cite{Buchdahl} with or without the LHY corrections.

\smallskip

Moreover, the inclusion of the LHY correction shifts the M-R profile upwards. For a given stellar mass the correction implies a larger radius. Consequently the inclusion of the LHY correction reduces the factor of compactness of condensate dark stars, which is illustrated in the lower panel of Figure \ref{fig:1} and \ref{fig:3}.

\smallskip

Next, the Harrison-Zeldovich Novikov criterion \cite{Harrison, ZN} states that the condition
\begin{equation}
\frac{dM}{d \rho_c} > 0
\end{equation}
indicates stable configurations, whereas the condition
\begin{equation}
\frac{dM}{d \rho_c} < 0
\end{equation}
indicates unstable configurations. In the upper panel of Figures \ref{fig:2} and \ref{fig:4} the stellar mass increases with the central energy density until it reaches a highest mass, and after that $M$ monotonically decreases with $\rho_c$. In the first part of the curve the derivative is positive, and therefore according to the HZN criterion the configurations are stable, whereas the other part of the curves corresponds to unstable configurations. The inclusion of the LHY corrections shifts the curves upwards, and so for a given central energy density, the correction implies a more massive star.

Finally, as shown in the lower panel of Figures \ref{fig:2} and \ref{fig:4}, the inclusion of the LHY correction increases the tidal Love number for a given factor of compactness. The impact of the correction is small in the case of model A, and more significant in the case of model B, especially at small factor of compactness.

Before we summarize our work and finish with the concluding remarks, let us add in passing that regarding future work it would be both interesting and timely to study radial and non-radial oscillations, since the frequencies of pulsating stars are very sensitive to inner structure and composition, and Asteroseismology has been proven to be a powerful tool to probe star interiors. What is more, we would like to check whether or not the Lee-Huang-Yang correction upsets universal relations \cite{Yagi:2013bca, Yagi:2013awa}. Finally, apart from quantum fluctuations the impact of thermal fluctuations should be investigated as well, see \cite{Gruber:2014mna}. Moreover, since rotation could be present in condensate dark stars, the inclusion of a non-vanishing angular momentum should have a considerable effect on structural properties and other observables of interest. We hope we will be able to address some of those issues in the future.


\section{Summary and Concluding Remarks}

To summarize our work, in the present article we have studied for the first time condensate dark stars beyond the mean field approximation taking into account the Lee-Huang-Yang correction. We have considered bosonic, self-interacting dark matter, which may solve the core/cusp problem, and we have studied structural properties of static, spherically symmetric self-gravitating configurations of a dilute and ultra-cold Bose gas of weakly interacting identical particles.

\smallskip

First we briefly reviewed hydrostatic equilibrium of fluid spheres within GR, and the structure equations that permit us to obtain interior solutions. Then we reviewed in some detail the Gross-Pitaevskii-Poisson equations for the coupled system of gravitational potential to the wave-function of the BEC. Within the mean-field approximation (or equivalently the Hartree approximation) we derived the polytropic equation of state that has been used extensively in studies of condensate dark stars. Finally, we incorporated the Lee-Huang-Yang correction that takes into account quantum fluctuations.

\smallskip

We have considered two models, and we have integrated the TOV equations numerically imposing appropriate conditions at the center of the stars. We have made use of the matching conditions at the surface of the stars in order to compute the mass and the radius of condensate dark stars as well as the factor of compactness, $C=M/R$. We have generated plots showing for both models the stellar mass, the radius, the factor of compactness, tidal Love numbers and stability of the condensate dark stars with and without the Lee-Huang-Yang correction. Our results show that the correction has a significant impact on the structural properties of the objects. In particular, our results indicate that the inclusion of the correction implies an EoS that can accommodate a larger highest mass, while at the same time it reduces the factor of compactness, since for a given stellar mass it increases the radius of the objects. Moreover, the inclusion of the LHY correction increases the tidal Love number for a given factor of compactness. The impact of the correction is small in the case of model A, and more significant in the case of model B, especially at small factor of compactness. Finally, regarding stability based on the Harrison-Zeldovich-Novikov criterion, the inclusion of the correction implies more massive stars for a given central density.

\section{Appendix: Matsubara frequencies and Green's functions in the quantum many-body problem}

The quantum many-body problem, Green's functions in the Matsubara formalism \cite{Matsubara:1955ws}, and diagrammatic techniques, both in zero and non-zero temperatures, are discussed for instance in the standard textbooks \cite{Walecka, Mahan, Mattuck}.

\smallskip

We consider a weakly interacting boson system at very low temperature $0 < T < T_c$, with $T_c$ being the critical temperature of the BEC. We assume a short-range repulsive interaction of the form
\begin{equation}
V(\vec{r}_1,\vec{r}_2) = g \: \delta^3(\vec{r}_1-\vec{r}_2),  \; \; \; \; \; \; g>0
\end{equation}
where the positive coupling constant is given by
\begin{equation}
g = \frac{4 \pi a_s}{m}
\end{equation}
with $m$ being the mass of the particles in the dilute and ultracold Bose gas. This coupling implies a low-energy elastic cross-section $\sigma$ of hard sphere
\begin{equation}
\sigma = 4 \pi a_s^2
\end{equation}
with $a_s$ being the scattering length. Thanks to the Dirac delta function, the Fourier transform of the two-body potential is a constant, $\tilde{V}(q)=g$.

The free Green's function $G_0(i\omega_n, \vec{k})$ is given by
\begin{equation}
G_0(i\omega_n, \vec{k}) = \frac{1}{i \omega_n-\xi_k}
\end{equation}
where $\xi_k=\epsilon_k - \mu$, with $\mu$ being the chemical potential, while the bosonic Matsubara frequencies are given by
\begin{equation}
\omega_n = \frac{2 n \pi}{\beta}, \; \; \; \; \; \; \beta = \frac{1}{T}, \; \; \; \; \; \;  n=0, 1, 2, 3,...
\end{equation}
Recall that the chemical potential vanishes at the critical temperature and below, $\mu=0, \; \;T \leq T_c$.
Furthermore, the Dyson equation for the full Green's function reads
\begin{equation}
G(i\omega_n, \vec{k}) = \frac{1}{i \omega_n-\xi_k - \Sigma_k} = \frac{1}{i \omega_n-\epsilon_k - \Sigma_k},
\end{equation}
where $\Sigma_k$ is the self-energy, which may be computed applying the Feynman rules of the diagrammatic expansion, which are the following:

\begin{itemize}

\item Associate a factor $\tilde{V}(q)$ with each interaction line.

\item Associate a factor $G_0$ with each free boson line.

\item Conservation of energy-momentum with each vertex.

\item Associate a wave-number $\vec{k}$ and Matsubara frequency $\omega_n$ with each loop.

\item Sum over Matsubara frequencies 
\begin{equation}
\frac{1}{\beta} \sum_n f(i \omega_n)
\end{equation}

\item Sum over the wave vector $\vec{k}$
\begin{equation}
\sum_k  \rightarrow \int \frac{d^3\vec{k}}{(2 \pi)^3},
\end{equation}
where in the thermodynamic limit the sum is converted to integrals over $\vec{k}$, while the summations over Matsubara frequencies are performed with the help contour integrals \cite{Nieto:1993pr} making use of the theory of complex functions and the residue theorem.

\end{itemize}

According to the Hartree approximation (Feynman diagram (a) of the Figure below) the self-energy is given by
\begin{equation}
\Sigma_k = \frac{1}{\beta}\sum_{n,k} \tilde{V}(0) G_0(i\omega_n, \vec{k}) = \frac{1}{\beta}\sum_{n,k} g \frac{1}{i \omega_n-\xi_k}.
\end{equation}
The summation over Matsubara frequencies yields
\begin{equation}
\frac{1}{\beta}\sum_{n} \frac{1}{i \omega_n-\xi_k}  = - n_B(\xi_k)
\end{equation}
and finally 
\begin{equation}
\Sigma_k = -g \int \frac{d^3\vec{k}}{(2 \pi)^3} n_B(\xi_k) = - g n
\end{equation}
where $n$ is the number density of the boson system, and $n_B$ is the Bose-Einstein distribution
\begin{equation}
n_B = \frac{1}{e^{\beta \epsilon_k}-1}.
\end{equation}
Therefore, the full Green's function within the Hartree approximation is computed to be
\begin{equation}
G(i\omega_n, \vec{k}) = \frac{1}{i \omega_n-\epsilon_k + g n} =  \frac{1}{i \omega_n-(\epsilon_k - g n)},
\end{equation}
which implies that the constant self energy corresponds to a shift to the chemical potential, $\mu=g n$.

\vskip 1cm

\includegraphics[width=10cm]{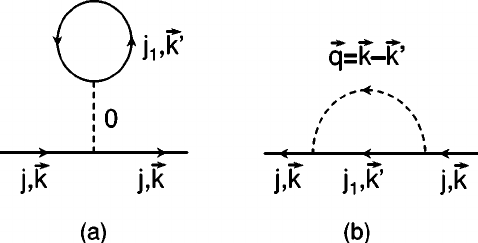}

\end{document}